\documentclass[%
 reprint,
 amsmath,amssymb,
 aps,
]{revtex4-2}
\usepackage{xcolor}

\usepackage{graphicx}
\usepackage{dcolumn}
\usepackage{bm}
\usepackage{hyperref}
\usepackage{multirow}
\usepackage{xspace}
\newcommand{\kT}{$k_T$\xspace}
\newcommand{\akT}{anti-$k_T$\xspace}
\newcommand{\pT}{$p_T$\xspace}
\newcommand{\pp}{$p$+$p$\xspace}
\newcommand{\Au}{Au+Au\xspace}
\newcommand{\Pb}{Pb+Pb\xspace}
\newcommand{\GeV}{GeV/$c$\xspace}
\newcommand{\sNN}{$\sqrt{s_{\mathrm{NN}}}$\xspace}
\newcommand{\sqrts}{$\sqrt{s}$\xspace}

\newcommand{\fref}[1]{Fig.~\ref{#1}}
\newcommand{\eref}[1]{eq.~\ref{#1}}

\newcommand{\Fref}[1]{Figure~\ref{#1}}

\newcommand{\pikp}{$\pi^{\pm}$, K$^{\pm}$, p and $\bar{p}$\xspace}
\newcommand{\BG}{TennGen\xspace}
\newcommand{\PYTHIA}{\textsc{PYTHIA}\xspace}
 
\begin{document}

% \preprint{APS/123-QED}

\title{Interpretable Machine Learning Methods\\ Applied to Jet Background Subtraction in Heavy Ion Collisions}
\author{Tanner Mengel} 
\author{Patrick Steffanic} 
\author{Charles Hughes}
 \author{Antonio Carlos Oliveira da Silva} 
\author{Christine Nattrass} 
\affiliation{University of Tennessee, Knoxville, TN, USA-37996.}

\url{https://doi.org/10.1103/PhysRevC.108.L021901}
\date{\today}

\begin{abstract}
Jet measurements in heavy ion collisions can provide constraints on the properties of the quark gluon plasma, but the kinematic reach is limited by a large, fluctuating background. We present a novel application of symbolic regression to extract a functional representation of a deep neural network trained to subtract background from jets in heavy ion collisions. We show that the deep neural network is approximately the same as a method using the particle multiplicity in a jet. This demonstrates that interpretable machine learning methods can provide insight into underlying physical processes.
\end{abstract}

\maketitle

\section{\label{sec:introduction}Introduction}
The Quark Gluon Plasma (QGP) is a hot, dense, strongly interacting liquid of quarks and gluons that is created briefly in high energy heavy ion collisions~\cite{Adcox:2004mh,Adams:2005dq,Back:2004je,Arsene:2004fa}. Measurements of jets produced by hard scatterings between partons in heavy ion collisions can be used to investigate the properties of the QGP~\cite{Connors:2017ptx}. Quantitative comparisons between jet measurements and physics models can provide further constraints on these properties~\cite{Burke:2013yra,JETSCAPE:2021ehl}. However, heavy ion events are dominated by a fluctuating background of soft particles not due to hard scatterings. The details of these fluctuations are sensitive to correlations from hydrodynamical flow and the shape of the single particle spectra~\cite{Hughes:2020lmo}, and as such are unlikely to be exactly the same in data and models. Mixed events are able to successfully describe the background in measurements of hadron-jet correlations by the STAR collaboration~\cite{STAR:2013thw} at the Relativistic Heavy Ion Collider (RHIC). Studies of the background at the Large Hadron Collider (LHC) by the ALICE Collaboration found that the distribution of background energy density in random cones is well described by a random background with correlations due to hydrodynamical flow and Poissonian fluctuations~\cite{ALICE:2012nbx}. A better understanding of this background will facilitate more precise jet measurements for comparisons between data and models.

Measurement precision and kinematic range is limited by the ability to correct for this background and its fluctuations. Background correction in jet measurements requires subtraction of contributions from soft particles within the jet, and suppression of fluctuations which have been reconstructed as combinatorial jets. At low momenta, combinatorial jets limit the kinematic reach of the measurement. Improved background subtraction methods would increase measurements' sensitivity to partonic energy loss. Measurements of jet spectra which extend to low momenta primarily use the area method~\cite{Soyez:2009cw} for background subtraction. This method was initially proposed to correct for the underlying event in \pp collisions in high pile-up conditions~\cite{Soyez:2009cw} and has also been applied to heavy ion collisions~\cite{ALICE:2013dpt,ALICE:2015mjv,ALICE:2019qyj,STAR:2020xiv}.

The complexity of jet background subtraction makes it an interesting environment to apply machine learning techniques. However, application of machine learning methods to background subtraction should be handled with care since models are not able to fully reproduce background fluctuations in heavy ion collisions~\cite{Hughes:2020lmo}. Nuclear physics has prioritized the continued advancement in machine learning analysis techniques with a focus on interpretable methods that are robust, provide clear uncertainty quantification, and are explainable~\cite{Achenbach:2023pba}. Applications of non-interpretable machine learning methods are insufficient when models available for training may be inaccurate, when it may be necessary to understand the method to interpret the results, or when a result is needed outside of the training space. 

Application of a deep neural network, i.e. a neural network with multiple hidden layers, to jet background subtraction in heavy ion collisions has demonstrated significant improvements compared to the area method, particularly at low jet momenta~\cite{Haake:2018hqn,ALICE:2023waz}. Deep neural networks are susceptible to model bias because their predictions risk being unreliable outside the domain of their training space. These methods may break down when they are extrapolated beyond this space, and due to their opaque nature, offer little indication where and why this break down occurs. In addition, one cannot validate the technique against data because we do not know the true jet momenta in data.

Increased performance of machine learning methods over traditional methods is an indication that there is information accessible to the machine learning that accounts for this improvement. We present an interpretable machine learning technique that allows us to understand why a deep neural network improves the jet momentum resolution in heavy ion collisions. We empirically derive an alternate method based on the background described in~\cite{TANNENBAUM200129,ALICE:2012nbx}, we call the multiplicity method. We compare the widths of the fluctuations of the jet momenta for the this method to the area and neural network methods and estimate the impact of the methods on the kinematic range. We apply symbolic regression to determine a functional form describing the mapping learned by the neural network, which was trained using \BG~\cite{Hughes:2020lmo} for the background and \PYTHIA~\cite{Sjostrand:2007gs} for the signal. We compare this functional description of the neural network to the form of the multiplicity method.

\section{\label{sec:method}Method}
\subsection{\label{sec:simulation}Simulation}
\BG~\cite{TennGen,TennGen200} generates heavy ion collisions with \pikp hadrons with yields~\cite{Aamodt:1313050}, momentum distributions~\cite{PHENIX:2013kod,Abelev:2013vea}, and azimuthal anisotropies~\cite{PHENIX:2014uik,Adam:2016nfo} matched to published data. \BG was updated to simulate collision energies per nucleon of \sNN = 200 GeV collisions as well as \sNN = 2.76 TeV, including multiplicity fluctuations, and improved computational efficiency.
Proton-proton collisions at \sqrts = 200 GeV were simulated with the \PYTHIA 8.307~\cite{Sjostrand:2007gs} Monash 2013 tune~\cite{Skands:2014pea} in 25 $p_{T}^{hard}$ bins starting at 5 GeV, with 1 million \pp events in each bin. Only final state charged particles from PYTHIA are mixed with a \BG background event. Charged particles from both \PYTHIA and \BG are required to have a minimum \pT of 150 MeV and be within pseudo-rapidity $|\eta| <$ 0.9.

Jets are clustered using the \akT algorithm with
FastJet 3.4.0~\cite{Cacciari:2011ma} with jet resolution parameters $R$ = 0.2, 0.4, and 0.6. To determine the true momentum, jets are reconstructed separately in both \PYTHIA and the combined event. Jets in the combined \PYTHIA and \BG event are geometrically matched to a \PYTHIA jet if $\Delta R = \sqrt{\Delta\eta^{2} + \Delta\phi^{2}} < 0.1$ where $\Delta\eta$ and $\Delta\phi$ are the differences in $\eta$ and $\phi$ between the jets and there is a bijective match. Reconstructed jets are required to have \pT $>$ 5 GeV and be within pseudo-rapidity $|\eta_{jet}| < 0.9 - R$.  The momentum of the \PYTHIA jet is taken as the truth momentum, $p_{T,Jet}^{Truth} \equiv p_{T,Jet}^{PYTHIA}$.

\subsection{\label{sec:area}Area and multiplicity methods}
For area-based background subtraction, the jet area~\cite{Cacciari:2008gn} is estimating through the use of ``ghost" particles, jets are reconstructed using the \kT jet finder~\cite{Ellis:1993tq}. The corrected jet momentum is then estimated as
\begin{equation}
  p_{T,Jet}^{Corr, A} = p_{T,Jet}^{tot} - \rho A \label{Eq:pTArea},
\end{equation}\noindent where $A$ is the jet area, $\rho$ is the background momentum density per unit area, and $p_{T,Jet}^{tot}$ is the total momentum in the jet. The $\rho$ in an event is approximated as the median $p_{T,Jet}^{tot}/A$ for \kT jets because \kT jets are dominated by background.

To a good approximation, the standard deviation of the momentum residual $\delta p_T = p_T^{Corr} - p_T^{Truth}$  with the area method is given by
\begin{equation}
  \sigma_{\delta p_{T}} = \sqrt{N \sigma^{2}_{p_{T}} + (N + 2 N^2 \sum_{n=1}^{\infty} v_n^2) \langle p_{T} \rangle^{2}}
\label{eq:DeltaPtwidths_flow}
\end{equation}
\noindent where $N$ is the number of background particles in the jet, $\sigma_{p_{T}}$ is the standard deviation of the single track momentum distribution, $v_n$ are the coefficients of the azimuthal anisotropies of the single particle distributions, and $\langle p_T \rangle $ is the average momentum of background particles~\cite{ALICE:2012nbx}. This is derived by assuming each of the $N$ particles is drawn from a single track momentum distribution which is approximately a Gamma distribution, giving rise to the first term~\cite{TANNENBAUM200129}. The second term is from Poissonian fluctuations in the number of background particles and the third term is from fluctuations in the number of particles due to hydrodynamical flow. Deviations of the single track momentum distribution from a Gamma distribution and momentum dependence of the $v_n$ lead to slightly larger widths~\cite{Hughes:2020lmo}.

The area method is usually used instead of iterative background subtraction methods~\cite{Hanks:2012wv,CMS:2016uxf,ATLAS:2012tjt} for measurements of jets at lower momenta.  Iterative methods may suppress the fluctuations described in \eref{eq:DeltaPtwidths_flow} by estimating the local background and suppress combinatorial jets by requiring high momentum or energy constituents.  At low momenta, these requirements may impose a bias on the surviving jets.  Fluctuations and the contribution from combinatorial jets are generally higher with the area method, but with less bias.

We propose a multiplicity-based method as an alternative to the area method
\begin{equation}
  p_{T,Jet}^{Corr, N} = p_{T,Jet}^{tot} - \rho_{Mult}(N_{tot}- N_{signal}) , \label{Eq:pTN}
\end{equation}
\noindent where $N_{tot}$ is the total number of particles in the jet, $N_{signal}$ is the number of particles in the signal, and $N = N_{tot} - N_{signal}$. This leverages the fact that the natural variable in the background fluctuations is the number of particles, largely eliminating the second and third terms in \eref{eq:DeltaPtwidths_flow}.
The $\rho_{Mult}$ in an event is the mean transverse momentum per background particle, which is approximated as the median $p_{T,Jet}^{tot}/N_{tot}$ for \kT jets. 
 $N_{signal}$ is roughly described by models~\cite{ALICE:2014dla} and therefore can be estimated. 
 % This estimate can be improved using measured fragmentation functions in heavy ion collisions. 
 Measurements of $\gamma-h$ correlations~\cite{PHENIX:2020alr} and reconstructed jets~\cite{CMS:2014jjt,ATLAS:2018bvp} indicate that there are around 0.5 additional particles for $p_{T}^{jet} \approx 10$ \GeV and 1.0 additional particles  for $p_{T}^{jet} \approx 100$ \GeV in heavy ion collisions.  
 % constrain modifications for low energy jets and measurements using reconstructed jets constrain modifications for high energy jets~\cite{CMS:2014jjt,ATLAS:2018bvp}. 
 % LHC~\cite{PHENIX:2020alr,CMS:2014jjt,ATLAS:2018bvp}  using measurements of $\gamma-$jet fragmentation functions in \sNN = 200 \GeV \Au collisions~\cite{PHENIX:2020alr} and inclusive jet fragmentation functions in \sNN = 5.02 TeV \Pb collisions~\cite{CMS:2014jjt,ATLAS:2018bvp}, which provide constraints on medium modification to jet multiplicity at both extremes of jet momenta. These measurements show that there is less than a 1 particle enhancement for jets momenta $p_{T}^{jet}>100$ \GeV and a $\pm$0.5 particle modification for low momenta jets $p_{T}^{jet} = 10$ \GeV. 
 If this were applied as an additional uncertainty, it would be proportional to $\sigma_{N_{signal}}\cdot\rho_{Mult}$, or around 0.25--0.5 \GeV. 
 % Based on conservative estimates from measurements it is unlikely that there are more than $\pm 2$ additional particles from medium modifications at in any kinematic regime. In \sNN = 200 \GeV \Au collisions $\rho_{Mult} \approx 0.5$ \GeV, yielding a maximum uncertainty around $1.0$ GeV. This additional uncertainty term is not fixture of the multiplicity method and can easily be further constrained by additional measurements of medium modification.
 
 % Measurements of $\gamma-$jet fragmentation functions~\cite{PHENIX:2020alr} in \sNN = 200 GeV \Au collisions and inclusive jet fragmentation functions from \sNN \pb  This method would introduce an additional systematic uncertainty. In measurements of fragmentation functions~\cite{CMS:2014jjt,ATLAS:2018bvp}, there are at most 0.5 additional particles at jet momenta $p_{T}^{Jet}>100$ \GeV.  
 % Using a extremely conservative estimate, it is unlikely that there are more than 10 additional particles in any kinematic regime.  
 % The additional uncertainty in the jet energy introduced by the multiplicity method is proportional to $\sigma_{N_{signal}}\cdot\rho_{Mult}$.   If we conservatively estimate that there are $0.5 \pm 0.5$ additional particles, since $\rho_{Mult} \approx 0.5$ \GeV, this uncertainty would be around $0.5$ GeV. This is small compared to most experimental jet momentum resolutions of 10-20${\%}$. 
 % The standard deviation of the momentum residual in \eref{eq:DeltaPtwidths_flow} is expected smaller for the multiplicity method than the area method for 0--80\% central collisions.

\subsection{\label{sec:machinelearning}Machine learning methods}
A sufficiently complex neural network can interpolate any function, at the cost of transparency to the user. This poses an obstacle to application of deep neural networks in physics where understanding predictions and identifying their potential biases is crucial. Our approach to addressing this challenge is through symbolic regression, one example of interpretable machine learning, to extract mathematical expressions from trained deep neural networks. 
The resulting equations provide an effective description of the neural network's mapping between the input and output. By constraining the types of operations available, we can impose complexity and smoothness requirements. 

We train a deep neural network to predict the corrected jet momentum from the following input features: the uncorrected jet momentum, jet area, jet angularity, number of jet constituents, and seven leading constituent momenta. The architecture and input features of the network are motivated by previous application of neural networks to proton-proton jets with a thermal background~\cite{Haake:2018hqn}. The deep neural network is implemented with TensorFlow 2.10.0~\cite{tensorflow2015-whitepaper}. The deep neural network has three hidden layers consisting of 100, 100 and 50 nodes, each activated by a rectified linear unit (ReLU)~\cite{GoodBengCour16} function. The model is optimized with ADAM~\cite{GoodBengCour16} and the loss function is a modified mean squared error
\begin{equation}\label{Eqn:loss}
  \mathcal{L} = \langle ||p_{T,Jet}^{Truth} - p_{T,Jet}^{DNN}||^2 \rangle +\lambda\sum_{l=1}^{L}||\mathbf{W}_{l}||^{2},
\end{equation}
where $p_{T,Jet}^{DNN}$ is the predicted jet momentum, $p_{T,Jet}^{Truth}$ is the truth momentum, the last term is an $L^2(\lambda)$ regularization where $\lambda = 0.001$, $\mathbf{W}_{l}$ is the weight matrix of layer $l$, and the sum is over the $L$ layers. The regularization term penalizes redundancy and encourages sparsity in the final trained network. The network is trained using 50\% of the simulated jets while the remaining 50\% are reserved for testing.

Once the neural network is trained, it represents an approximate mapping between the input jet features and the truth jet momentum. 
We apply a genetic algorithm to symbolically regress a functional form which describes this mapping using the PySR 0.11.11~\cite{pysr} package. The PySR model samples the phase space of analytic expressions defined by operators, input features, and constants for minimization through genetic programming. The input features are comparable to those of the neural network, and the pool of operations are arithmetic, exponential, trigonometric, and exponentiation. The model mutates over 50 generations of 20 different population samples, with each population containing 33 individuals. The loss function for the PySR model
\begin{equation}\label{Eqn:lossPysr}
  \mathcal{L} = \langle ||p_{T,Jet}^{DNN} - p_{T,Jet}^{PySR}||^{2} \rangle,
\end{equation}
is the mean squared error between the prediction from PySR $p_{T,Jet}^{PySR}$ and the corrected jet momentum predicted by the neural network. PySR evaluates expressions based on a score $S$ that rewards minimizing the loss function $\mathcal{L}$ and penalizes equation complexity $C$
\begin{equation}
  S = -\frac{\delta \ln\mathcal{L}}{\delta C}, 
\end{equation}
where the equation complexity $C$ is defined as the total number of operations, variables, and constants used in an equation~\cite{cranmer2020discovering}. The simulated jets, designated for testing, are used to sufficiently sample the neural network outputs throughout the possible input feature space. The highest scoring PySR expression is a functional representation of the mapping from input jet features to corrected jet momentum learned by the deep neural network. 

\subsection{Unfolding}
% The improvement in jet momentum resolution extends the kinematic range of the measurement to lower jet momenta. Increased precision in jet momentum allows for combinatorial jets to be reconstructed in jet momentum bins closer to zero. 
The lower threshold for unfolding is typically set to be between 2-5 times the width of the jet momentum resolution to suppress effects of combinatorial jets on the unfolded results~\cite{ALICE:2013dpt,ALICE:2023waz}. We unfold the reconstructed jet momentum spectra using five iterations of the Bayesian unfolding method~\cite{DAGOSTINI1995487} in RooUnfold 2.0.0~\cite{Brenner:2019lmf}.
% Five iterations is when the change in $\chi^2$ between the unfolded and truth spectra becomes less than the uncertainties of the measured spectra. 
We construct a response matrix using \PYTHIA jets (truth jets) matched to \PYTHIA+\BG jets (reconstructed jets).
The momentum of the \PYTHIA jet is taken as the truth momentum, $p_{T,jet}^{Truth} \equiv p_{T,ket}^{PYTHIA}$.
We then unfold our reconstructed jet spectra. The reconstructed spectra has no matching criteria between the \PYTHIA+\BG jets and \PYTHIA jets and no kinematic cuts to suppress combinatorial jet contributions. We use reconstructed jet spectra including combinatorial background to investigate the sensitivity of the lower momentum threshold to combinatorial background.
% , since most of the combinatorial jets populate this kinematic region. We find that the width of the resolution does not impact the uncertainties on the unfolded spectra. The suppression of combinatorial jets for each method is demonstrated by taking the ratio of unmatched reconstructed jet spectra including combinatorial background to matched reconstructed \PYTHIA jet spectra.

\section{\label{sec:results}Results}
\begin{figure*}
  \centering
  \includegraphics[width=17.6cm,height=7.3cm]{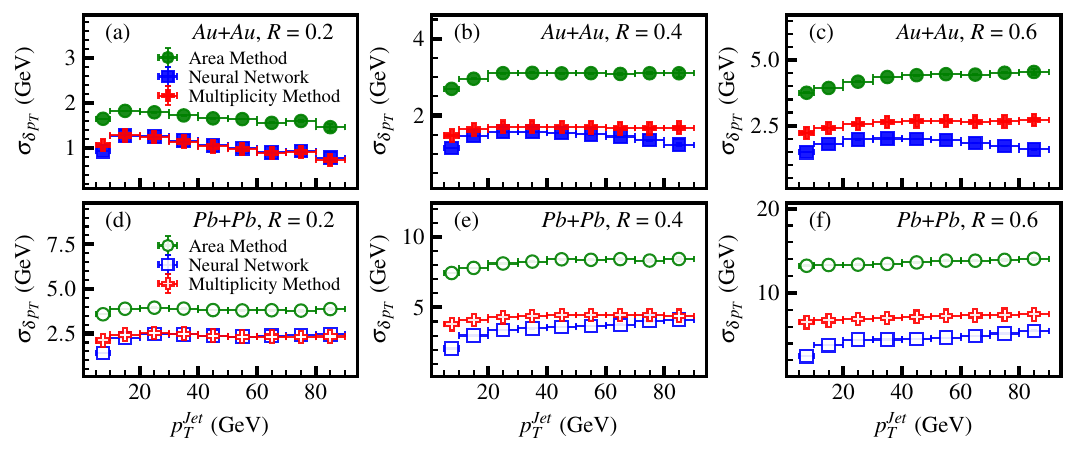}
  \caption{Comparisons of jet $p_{T}$ residual width for each background subtraction method as a function of reconstructed jet momentum for \Au collisions at \sNN = 200 GeV and \Pb collisions at \sNN = 2.76 TeV for jet resolution parameters $R=0.2, 0.4,$ and $0.6$.
  }
  \label{fig:ReconstructionWidthvspT}
\end{figure*}

\begin{figure}
  \centering
  \includegraphics[width= 8.6cm,height=10.6cm]{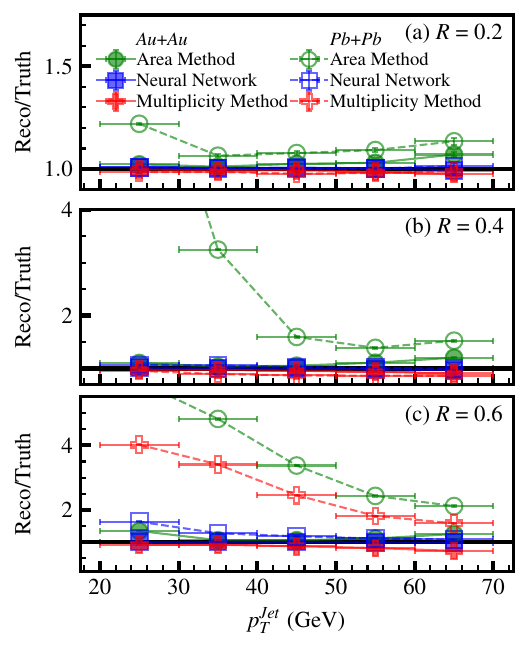}
  \caption{Ratio of the reconstructed jet spectrum over the truth spectrum for \Au collisions at \sNN = 200 GeV and \Pb collisions at \sNN = 2.76 TeV for jet resolution parameters $R=0.2, 0.4,$ and $0.6$. Low momentum points for the area method for LHC energies at $R$ = 0.4 and $R$ = 0.6 are off scale.  
  }
  \label{fig:UncorrectedJetRatio}
\end{figure}

\Fref{fig:ReconstructionWidthvspT} shows the width of the jet momentum residual distributions as a function of jet momentum for each background subtraction method in both \Au collisions at \sNN = 200 GeV and \Pb collisions at \sNN = 2.76 TeV. The $\sigma_{\delta p_{T}}$ increases with increasing jet resolution parameter, as expected because there is more background when the jet is larger. The $\sigma_{\delta p_{T}}$ also increases with \sNN because the particle multiplicity increases. As seen in~\cite{Haake:2018hqn}, the deep neural network reconstructs the momentum considerably more accurately than the area method. The performance of the multiplicity method is comparable to that of the deep neural network in \Au collisions and small jet resolution parameter. 
The ability of each method to sufficiently suppress contributions from combinatorial jets at low \pT is demonstrated with the ratios of the reconstructed jet spectra to the true jet spectra, shown in~\fref{fig:UncorrectedJetRatio}. The contributions from combinatorial jets decreases with increasing jet momentum for all methods, with all jet resolution parameters, and for both collision energies. The ratios for the deep neural network and multiplicity methods are both lower than those of the area method.

\begin{figure}
    \centering
    \includegraphics[width= 8.6cm,height=10.6cm]{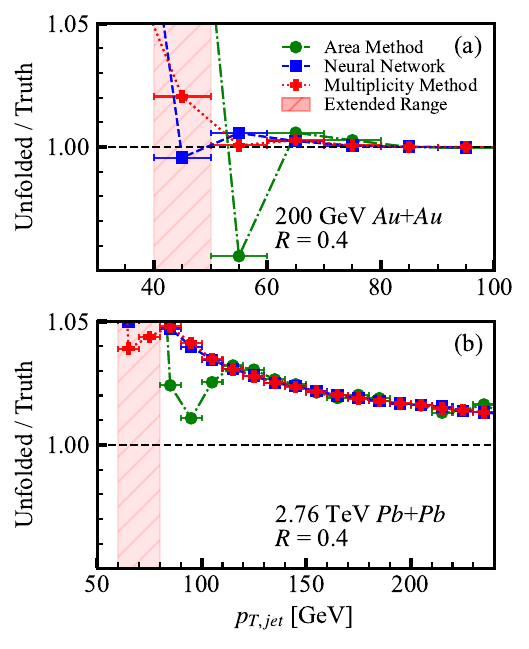}
    \caption{Ratio of unfolded jet spectrum over truth spectrum for (a) \Au collisions at \sNN = 200 GeV and (b) \Pb collisions at \sNN = 2.76 TeV for jet resolution parameter $R=0.4$.}
    \label{fig:UnfoldedJetRatio}
\end{figure}

The ratios of the unfolded spectra to the true jet spectra
% for jets with jet parameter R = 0.4 in \Au and \Pb collisions 
are shown in~\fref{fig:UnfoldedJetRatio}. Fluctuations from one at lower jet momenta are where the method becomes unstable due to overwhelming contributions from combinatorial background. Reconstructed jet spectra have no kinematic cuts to suppress combinatorial jet contributions therefore any extension in the lower kinematic range is due to the momentum resolution of the background subtraction method. 
% The shaded regions show where the unfolded spectra have less than a $5\%$ bias. The multiplicity and deep neural network methods extend the lower kinematic range by 8.5 GeV in \Au collisions and 16.5 GeV in \Pb collisions. 
The multiplicity and deep neural network methods are stable to at least 10 \GeV lower in momentum than the area method for all jet resolution parameters and collisions systems. 
% Table \ref{tab:KinematicRanges} shows the increase in lower kinematic reach for the multiplicity and deep neural network methods for all jet samples.  

% \begin{table}
%     \centering
%     \begin{tabular}{|c|c|c|}
%     \toprule
%      Coll. Sys. & Res. Par. (R) & Range Extension (\GeV)\\
%     \hline \hline 
%      \multirow{3}{*}{\Au}    &  0.2  & 10.0  \\
%      % \hline
%          &  0.4  &  10.0 \\
%      % \hline
%          &  0.6  &  10.0 \\
%      \hline
%      \multirow{3}{*}{\Pb}    &   0.2 &  10.0 \\
%      % \hline
%          &   0.4 &  20.0 \\
%      % \hline
%          &  0.6  &  10.0 \\
%      \hline
%      \hline
%     \end{tabular}
%     \caption{Extensions of the lower kinematic bound for unfolded jet spectra when using the multiplicity method compared to using the area method for various jet resolution parameters for  \sNN = 200 GeV \Au and \sNN = 2.76 TeV \Pb collisions.}
%     \label{tab:KinematicRanges}

% \end{table}

\begin{figure}
  \centering
  \includegraphics[width= 8.6cm,height=10.6cm]{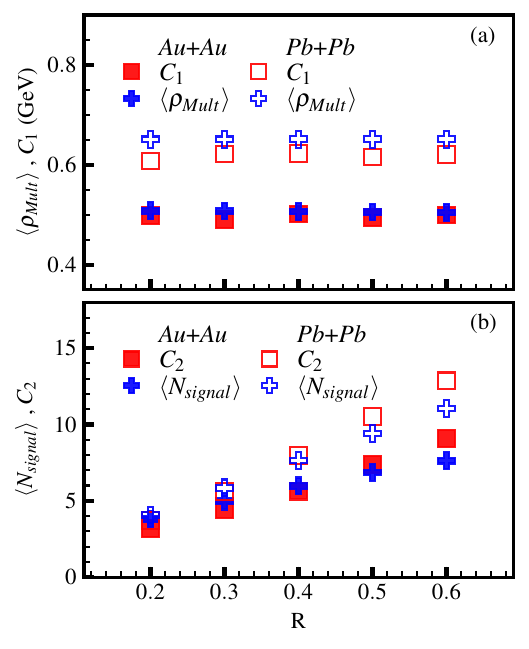}
  \caption{PySR optimization constants compared to average value of multiplicity method parameters versus jet resolution parameter for \Au collisions at \sNN = 200 GeV and \Pb collisions at \sNN = 2.76 TeV for jet resolution parameters $R=0.2, 0.4,$ and $0.6$. 
  }
  \label{fig:pysr_params}
\end{figure}

For all jet resolution parameters and both collision energies, the symbolic regression found that the best description of the deep neural network has the functional form
\begin{equation}
    p_{T,Jet}^{Corr. PySR} = p_{T,Jet}^{tot} - C_{1} (N_{tot}- C_{2}) \label{Eq:pTpysr},
\end{equation}
where the two parameters, $C_{1}$ and $C_{2}$, are optimization constants defined by PySR.
These parameters are plotted in \fref{fig:pysr_params} and compared to the average value of the parameters used in the multiplicity method. We find that the symbolic regression parameters $C_{1}$ and $C_{2}$ are comparable to the averages of those for the multiplicity method, $\langle \rho_{Mult} \rangle $ and $\langle N_{signal}\rangle $, respectively, with greater deviations at LHC energies and larger $R$. This indicates that the deep neural network is using a relationship similar to the multiplicity method to predict jet momenta. 
This approach to machine learning enables use of domain knowledge. The optimization parameters from PySR would otherwise not have a clear physical interpretation. Since these parameters are understood in the multiplicity method, it is possible to assign a physically motivated uncertainty to them.  Assumptions inherent in the method can then be understood.

\section{\label{sec:conclusions}Conclusions}
We have shown that interpretable machine learning methods can learn an underlying physical correlation, such as the multiplicity dependence for jet background, that was previously overlooked. The dependence of the neural network on jet multiplicity, rather than other input features, is easily understood since the fluctuations in the background are well described by eq.~\ref{eq:DeltaPtwidths_flow}~\cite{TANNENBAUM200129,ALICE:2012nbx,Hughes:2020lmo}, where the multiplicity is the dominant variable in the standard deviation. Using the multiplicity instead of the the area removes the second two terms in the standard deviation because these arise purely from fluctuations in the number of particles. 
The systematic uncertainty on $N_{signal}$ in the multiplicity method can be constrained by existing measurements and is therefore data-driven.  
% becomes model independent through a data driven approach for estimating $N_{signal}$ from measurements.
The multiplicity method achieves similar performance without the model dependence of the neural network.  

We previously showed that when we used a random forest to classify jets as either combinatorial or signal, the optimal selection was on the leading hadron momentum~\cite{Steffanic:2023cyx}, already used as a standard technique~\cite{ALICE:2013dpt,ALICE:2015mjv,ALICE:2019qyj,STAR:2020xiv}. We argue that applying machine learning to scientific problems requires methods that are interpretable. The definition of interpretability is often ambiguous or under-specified, but~\cite{lipton2018mythos} presents several definitions of interpretability to guide our selection of machine learning methods. We argue that for a machine learning method to be interpretable (1) it should be applicable equivalently to data and simulation, (2) the method's output can be understood outside the range of the training set, and (3) a measurement uncertainty can be calculated. We argue that an uncertainty on the method is not a proxy for a measurement uncertainty. These stricter criteria are consistent with those outlined in~\cite{Achenbach:2023pba}.
Symbolic regression satisfies these requirements because the output is a formula. The convergence of the empirically-based multiplicity method and the formula produced through symbolic regression is a clear indication of the usefulness of an interpretable method. Machine learning should be used to gain knowledge about the underlying physical processes that drive the relationships in our data. We must interpret the details of any method in terms of these underlying physical processes.

 \begin{acknowledgments}
We are grateful to Friederike Bock, Hannah Bossi, Adrian Del Maestro, Jamie Nagle, Ken Read, and Austin Schmier for useful discussions and feedback on the manuscript. This work was supported in part by funding from the Division of Nuclear Physics of the U.S. Department of Energy under Grant No. DE-FG02-96ER40982. This work was performed on the computational resources at the Infrastructure for Scientific Applications and Advanced Computing (ISAAC) supported by the University of Tennessee.
\end{acknowledgments}

\bibliography{prlbib}

\end{document}